\documentstyle[aps,twocolumn]{revtex}

\input{epsf}

\title{Quantum clock synchronization and quantum error correction}
\author{John Preskill\thanks{\tt
preskill@theory.caltech.edu}}

\address{Institute for Quantum Information\\
California Institute of Technology, 
Pasadena, CA 91125, USA }

\begin{document}

\maketitle

\begin{abstract}
I consider quantum protocols for clock synchronization, and investigate in particular whether entanglement distillation or quantum error-correcting codes can improve the robustness of these protocols. I also draw attention to some unanswered questions about the relativistic theory of quantum measurement. This paper is based on a talk given at the NASA-DoD Workshop on Quantum Information and Clock Synchronization for Space Applications (QuICSSA), September 25-26, 2000.
\end{abstract}

\section{Introduction}
The theme of this workshop, clock synchronization, has occupied a prominent position at the frontier of technology for a long time. From a recent article by Peter Galison \cite{galison}, I  learned that the International Conference on Chronometry in 1900 included a session devoted to the problem of clock synchronization. This was an urgent problem at the time, especially in Europe, where a single track would often carry railroad traffic in both directions, so that precise scheduling was necessary to avoid  disastrous collisions.

According to Galison, the technical community's keen interest in clock coordination 100 years ago could not have escaped the attention of a certain patent clerk in Bern named Einstein \cite{galison}:

\begin{quote}
Meanwhile, all around him, literally, was the burgeoning fascination with electrocoordinated time. Every day Einstein took the short stroll from his house, left down the Kramgasse to the patent office; every day he must have seen the great clock towers that presided over Bern with their coordinated clocks, and the myriad of street clocks branched proudly to the central telegraph office.
\end{quote}

\noindent To Galison, whose work often highlights the role of technology in the development of scientific ideas, it is irresistible to speculate that the preoccupation with clock coordination circa 1900 helped to steer Einstein to the insight that simultaneity is the key concept for understanding the electrodynamics of moving bodies, and so may have inspired the most famous scientific paper of the past century.

To me as a theoretical physicist, a pleasing outcome of a workshop like this one would be that our musings about clock synchronization lead to conceptual insights into the properties of quantum information. Some might hope for a flow of ideas in the other direction, but both directions can be beneficial.

\section{Two protocols}
\begin{figure}
\begin{center}
\leavevmode
\epsfxsize=3in
\epsfbox{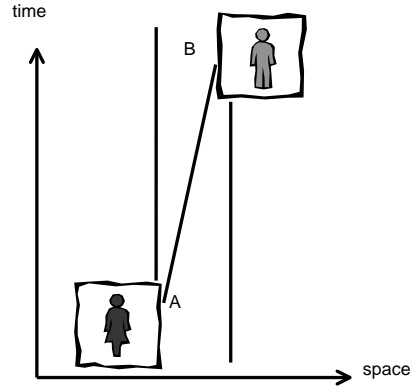}
\end{center}
\caption{Slow Clock Transport. Alice prepares precessing qubits, and sends them to Bob, who measures them. Bob can estimate the proper time that elapsed, between Alice's preparation and Bob's measurement, along the world line of the qubits.}
\label{fig:sct}
\end{figure}

Most of this talk will concern a scenario in which two parties, Alice and Bob, both have good local clocks that are stable and accurate, and wish to synchronize these clocks in their common rest frame. One method that works, and does not require Alice and Bob to have accurate knowledge of the distance between them, is Slow Clock Transport (SCT), illustrated in Fig.~\ref{fig:sct}. Alice has a traveling clock that she synchronizes with her local clock (event $A$) and then  sends to Bob, who receives it and reads it (event $B$). When Bob reads the clock it has advanced by $\tau_{AB}$, the proper time along the clock's world line that has elapsed during transit. If the clock moved slowly, then $\tau_{AB}$ is close to the elapsed time as measured by Alice's clock during the transit, so that Bob can synchronize his clock with Alice's.

\begin{figure}
\begin{center}
\leavevmode
\epsfxsize=3in
\epsfbox{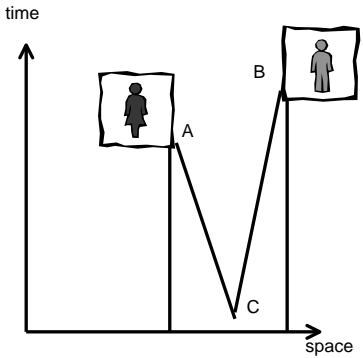}
\end{center}
\caption{Quantum Clock Synchronization. Charlie prepares entangled pairs of qubits, sends half of each pair to Alice, and sends half to Bob. Alice and Bob measure the qubits. After communicating classically, they estimate the proper time along the world line that travels backward from Alice's measurement to Charlie's preparation and then forward from Charlie's preparation to Bob's measurement.}
\label{fig:qcs}
\end{figure}

SCT works but it is not very sexy. It would be more fun to use a method that exploits the resource of quantum entanglement, what the JPL group \cite{jpl} called Quantum (Atomic) Clock Synchronization (QuACS or QCS), illustrated in Fig.~\ref{fig:qcs}. So now suppose that Alice and Bob have a co-conspirator Charlie, who prepares maximally entangled pairs (event $C$); let's suppose that the state of each pair is the singlet state $|\psi^-\rangle$. Charlie sends half of each pair to Alice, and half to Bob. Alice measures the observable $X=\pmatrix{0 & 1\cr 1 & 0}$ of her qubits (event $A$), and Bob measures X of his qubits (event $B$). Comparing their results, Alice and Bob can infer the value of
\begin{equation}
\tau_{BC} - \tau_{AC}~,
\end{equation}
the proper time along the world line of the qubits that moves from $A$ backward in time to $C$ and then forward in time to $B$. If the qubits were transported slowly, this difference of proper times is close to the time difference $t_A-t_B$ in the reference frame in which Alice and Bob are at rest.

Both protocols work, but QCS is more technically demanding than SCT, so why would we prefer QCS? Perhaps if we need to synchronize periodically, rather than just once, we'll find it convenient to ship many pairs of qubits ahead of time and save them until they are needed, rather than sending another clock on demand every time we need to synchronize. But when I first heard about the idea that quantum information could be used for clock synchronization (from Hideo Mabuchi \cite{hideo}), what seemed intriguing to me is that we might be able to correct phase errors that afflict the traveling qubits, by {\em purifying} the shared entanglement. In \S III-VII, I'll explain why I haven't been able to get this idea to work.
 
\section{QCS with flawed pairs}
To assess whether entanglement purification might improve the accuracy of clock synchronization, let us begin by looking at the QCS protocol in more detail. Our qubits are two-level atoms, each governed by the Hamiltonian $H={1\over 2}\omega Z$, $Z=\pmatrix{1&0\cr 0 & -1}$. The pairs prepared by Charlie are in the state
\begin{equation}
|\psi^-\rangle_{AB}=|01\rangle_{AB} - |10\rangle_{AB}~,
\end{equation}
which is stationary (I'm not bothering to write factors of $\sqrt{2}$ here.) When Alice measures $X$, obtaining the outcome $\pm 1$, she prepares for Bob on the same time slice the state $|\mp\rangle_B = |0\rangle_B \mp |1\rangle_B$, which evolves after time $t$ to $|0\rangle_B \mp e^{i\omega t}|1\rangle_B$. Then Bob measures $X$, obtaining the outcome $+1$ with probability 
\begin{equation}
P(+_B|\pm_A)={1\over 2}\left(1\mp \cos\omega t\right)~.
\end{equation}
Alice and Bob confer so that Bob knows for which qubits the result of her $X$ measurement was $+1$. Then with $n$ qubits, the time $t$ can be determined to an accuracy $\Delta t = \omega^{-1} n^{-1/2}.$

It is instructive to consider what would happen if the pairs prepared by Charlie were not really $|\psi^-\rangle$'s, but were instead in the state
\begin{eqnarray}
|\psi^-(\Delta)\rangle_{AB}&= &|01\rangle_{AB} - e^{i\omega \Delta}|10\rangle_{AB} = \nonumber\\
&=& I\otimes U_\Delta^{-1}|\psi^-\rangle_{AB}~, \quad U_\Delta=e^{-iH\Delta}~.
\end{eqnarray}
This is the state that would be obtained if the time evolution operator, for Bob's qubit only, were to act on the initial state $|\psi^-\rangle_{AB}$ for time $-\Delta$. With the pairs in this state, when Alice measures $X$ she prepares for Bob one of the states $U_\Delta^{-1}|\mp\rangle_B$, which will evolve in time $t$ to
\begin{equation}
U_tU_\Delta^{-1}|\mp\rangle_B= U_{t-\Delta}|\mp\rangle_B~.
\end{equation}
Therefore, the apparent time offset detected in the QCS protocol will be $t-\Delta$ rather than $t$. If, without telling Alice and Bob, Charlie replaces the $|\psi^-\rangle$'s by $|\psi^-(\Delta)\rangle$'s, then Alice and Bob will think that their clocks are synchronized even though Bob's really lags behind Alice's by $\Delta$.

\section{Phase error correction for clock synchronization?}
If Alice and Bob perform SCT or QCS, dephasing of the qubits will weaken the signal. If Bob measures his qubit a time $t$ after Alice measures $X=\pm 1$ of hers, the probability that Bob finds $X=1$ is
\begin{equation}
P(+)={1\over 2}\left(1\mp \eta \cos\omega t\right)~;
\end{equation}
here $\eta\le 1$ is the phase damping factor, {\it e.g.} $\eta=e^{-\Gamma T}$ where $T$ is the time that the qubits have been exposed to phase noise and $\Gamma^{-1}$ is the damping time. (We have assumed that there are no systematic phase errors -- the noise has zero mean.)

\begin{figure}
\begin{center}
\leavevmode
\epsfxsize=3.5in
\epsfbox{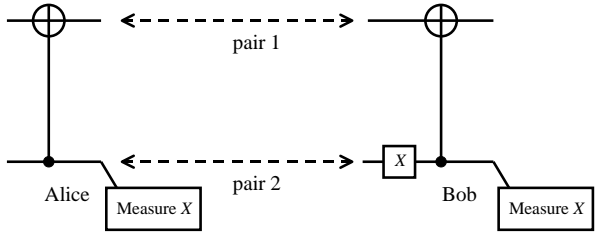}
\end{center}
\caption{Purification protocol that protects entangled pairs against phase errors. Alice and Bob compare the phase bits of pair 1 and pair 2 by destroying pair 2. If the phase bits disagree they discard pair 1 and if the phase bits agree they retain pair 1.}
\label{fig:purify}
\end{figure}

The damage to the qubits caused by phase damping might be reversed by an entanglement purification protocol, where an initial supply of noisy entangled pairs is ``distilled'' to a smaller number that approximate $|\psi^-\rangle$ with better fidelity \cite{bdsw}. Such a protocol is illustrated in Fig.~\ref{fig:purify}. Alice and Bob select two pairs, and each performs an operation on her/his half of the two pairs, culminating in a bilateral $X$ measurement that destroys one of the two pairs. If Alice and Bob get the same measurement result when they measure pair number 2, then they retain pair number 1; otherwise, they throw pair number 1 away.

How does it work? Alice and Bob want to have pairs with $X_A\otimes X_B=-1$. Their bilateral procedure allows them to measure 
\begin{equation}
\left(X_A\otimes X_B\right)_1\cdot \left(X_A\otimes X_B\right)_2=\pm 1~,
\end{equation}
The value is $-1$ if one of the pairs has a phase error, and it is $+1$ if either both pairs are good or both pairs are bad. Hence, as long as the original ensemble approximates $|\psi^-\rangle$ with good enough fidelity ({\it e.g.}, $F>1/2$), the pairs that are retained have higher fidelity than the original pairs.

We need to notice though, that in order for the purification protocol to achieve its intended purpose, Alice's and Bob's operations must be perfectly synchronized. If Bob's operations were to systematically lag behind Alice's by time $\Delta$, then the protocol would actually distill an eigenstate of 
\begin{equation}
X_A\otimes \left(U_\Delta^{-1}X_B U_\Delta\right)~,
\end{equation}
namely the state $|\psi^-(\Delta)\rangle= I\otimes U_\Delta^{-1}|\psi^-\rangle$.

Therefore, if Alice and Bob first distill their pairs and then perform QCS, all they will be able to detect is a {\em relative} offset, the difference between the offset used in the purification protocol and the offset used in QCS. But this is information they could discern by reading their accurate local clocks. They would not gain anything from consuming their shared entanglement.

\section{Phase stabilization through quantum error-correcting codes?}

We might try this alternative procedure: To combat dephasing, Charlie does not prepare raw entangled pairs of qubits, but instead he {\em encodes} the pairs using a quantum error-correcting code that resists phase errors. The encoded qubits can be actively stabilized as they travel to Alice and Bob, so that they are sure to arrive safely. Then Alice and Bob can execute QCS.

But if phase errors are the enemy then there is a problem, because the phase rotation due to the natural evolution of the raw qubits is one kind of error that the code is designed to resist. The stabilization of the encoded state freezes this natural evolution. Alice can decode her qubit and measure $X$, but if Bob is still preserving his encoded qubit, Alice's action will not ``start the clock'' of Bob's qubit on the same time slice; rather its evolution will remain frozen.

In fact then, all Alice and Bob can learn about if they first decode and then execute QCS, is a {\em relative} offset: the difference of the offset used in decoding and the offset of the final measurements. Again, this is information that can be inferred by referring to Alice's and Bob's local clocks.

The difficulty we have encountered here is closely related to the obstacle that has so far prevented us from finding a powerful way to use quantum error-correcting codes to improve frequency standards. We would like to use quantum error correction to stabilize a {\em precessing} qubit that can serve as an accurate standard; therefore the natural evolution of the qubits should preserve the code subspace ({\it i.e.}, should commute with the code stabilizer). This requirement strongly restricts the error-correcting power of the code.

One code with the desired property is the repetition code, which can correct coherent bit flip errors (that is, stochastic $X$ errors); the encoded state evolves as
\begin{equation}
|\psi(t\rangle= |000\dots 0\rangle + e^{in\omega t}|111\dots 1\rangle~.
\end{equation}
This is just the rapidly precessing cat state whose advantages have been extolled by the NIST group \cite{nist}. If the limiting factor in the accuracy of the standard were a ``bit-flip'' channel, then this state could be actively stabilized by quantum error correction.

\begin{figure}
\begin{center}
\leavevmode
\epsfxsize=3in
\epsfbox{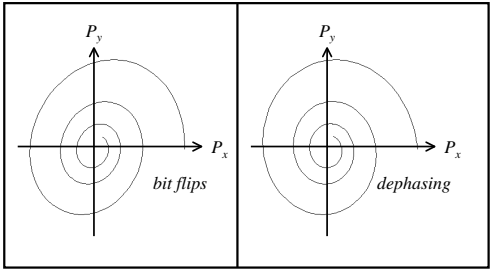}
\end{center}
\caption{Depolarizing spins. The time evolving polarization of a qubit, found by integrating the master equation, is shown for a bit flip channel (left) and a dephasing channel (right).}
\label{fig:master}
\end{figure}

As a mathematical statement, this is not completely vacuous, since it is possible in principle for the precision in the measurement of the precession rate to be limited by bit-flip errors.
Fig.~\ref{fig:master} shows the decay of the polarization of a qubit, found by integrating the master equation, for a bit-flip channel and a dephasing channel with comparable rates. Qualitatively, they are similar; the visible difference is that the bit-flips ($X$ errors) do not damage the $x$-component of the polarization. The bit-flip damage can be controlled by the quantum repetition code; the phase damping cannot be. 

While correct as a mathematical statement, this is not a very useful observation, since I don't know of any realistic physical setting in which a bit-flip channel really limits the accuracy of an interesting measurement.

\section{Quantum or classical?}
What is really ``quantum'' about the QCS protocol? Of course, each ``clock'' is a qubit, so when we read it we acquire at best one bit of information, and we need many qubits to estimate a time offset accurately. But it is worthwhile to emphasize that the correlations that were exploited might as well have been classical rather than quantum correlations. That is, although we have said that the QCS protocol  exploits entanglement, we have only used the property that the state $|\psi^-\rangle_{AB}$ has two qubits with correlated values of $X$; it makes no difference to us that the qubits are also correlated in the complementary variable $Z$. 

Therefore, it would work nearly as well for Charlie to prepare a product state $|+\rangle_A\otimes |-\rangle_B$. Unlike the entangled state $|\psi^-\rangle_{AB}$, this state is not stationary; in time $T$ it evolves to 
\begin{equation}
\left(|0\rangle_A + e^{i\omega T} |1\rangle_A\right)
\otimes \left(|0\rangle_B - e^{i\omega T} |1\rangle_B\right)~.
\end{equation}
If Alice and Bob both measure $X$, how strongly correlated their results are depends on the time offset of their measurements, as is the case if Charlie prepares $|\psi^-\rangle$'s.

Specifically, if there is no dephasing of the qubits, then if Bob measures a time $t$ after Alice, and we average over the time $T$ since Charlie's preparation (which is assumed to be completely unknown), then the probability that Bob finds $X=+1$ if Alice does is 
\begin{equation}
P(+_B|+_A;t)= {1\over 2}\left(1-{1\over 2}\cos\omega t\right)~.
\end{equation}
The signal is weaker then if Alice and Bob perform QCS, so that they need $4n$ product pairs to find $t$ to the same accuracy as with $n$ entangled pairs, but that is not a very heavy price to pay in return for the advantages of working with product states rather than entangled states.

(Of course, with SCT we don't have to pay even that price -- $n$ qubits sent from Alice to Bob are as effective as $n$ entangled pairs measured by Alice and Bob.) 

What about the purification? Since it involves coherent processing, there may seem to be something quantum about it --- I'm not sure I would know how to make a collective observation on two pairs of analog classical clocks without actually reading each clock. But since the correlations are essentially classical, we are not really making any use of the essentially quantum features of entanglement purification.

If we are really interested in, say, using our entanglement for EPR quantum key distribution, then purification may be essential for achieving {\em quantum privacy amplification} --- it is how we ensure that no potential eavesdropper has more than a negligible amount of information about the key. Since in QCS we are really exploiting a correlation that is essentially classical, the purification doesn't buy us anything, at least if there are no systematic phase errors (if the phase noise has zero mean). For example, suppose that our supply of pairs is a mixture of $|\psi^-\rangle$'s and $|\psi^+\rangle$'s with fidelity $F$:
\begin{equation}
\rho= F |\psi^-\rangle\langle \psi^-| + (1-F) |\psi^+\rangle\langle \psi ^+|~.
\end{equation}
Then Bob's measurement of his precessing qubit is governed by the probability distribution
\begin{equation}
P(+_B|+_A)= {1\over 2}\left( 1 \mp (2F-1) \cos\omega t\right)~,
\end{equation}
so that with $n$ pairs we can determine $t$ to accuracy
\begin{equation}
\Delta t= (2F-1)^{-1}\omega^{-1}n^{-1/2}~.
\end{equation}
If we perform one round of purification (perfectly synchronized), the number of surviving pairs is reduced (on average) to
\begin{equation}
n'= n\cdot {1\over 2}\left(F^2 + (1-F)^2\right)~,
\end{equation}
while the fidelity of the remaining pairs improves to
\begin{equation}
F' = {F^2\over {F^2 + (1-F)^2}}~.
\end{equation}
Therefore, if we perform QCS with the remaining pairs we can determine $t$ to the accuracy
\begin{eqnarray}
\Delta t'&=& \Delta t\cdot {2F-1\over 2F'-1}\cdot \sqrt{n\over n'}\nonumber\\
&=& \sqrt{2}\cdot \left(F^2 +(1-F)^2\right)^{1/2}~.
\end{eqnarray}
If the fidelity is close to one, purification hurts, because we waste half the pairs needlessly. Even for fidelity $F=1/2+\epsilon$ with $\epsilon \ll 1$, purification doesn't help: we can boost $\epsilon$ by a factor of about two, but we're no better off because the number of pairs is reduced by a factor of about 4.

In principle, purification might reduce {\em systematic} phase errors. For example, suppose that all of the pairs are identical, but with the unknown phase error $\delta$
\begin{equation}
|01\rangle - e^{i\delta}|10\rangle~.
\end{equation}
Though the states are maximally entangled, Alice and Bob can't fully exploit the entanglement if they don't know the value of $\delta$. Now if they execute the (perfectly synchronized) purification protocol, the number of pairs is reduced (on average) to
\begin{equation}
n'=n\cdot {1\over 2}\left[\cos^4(\delta/2) + \sin^4(\delta/2)\right]~, 
\end{equation}
and the phase of the remaining pairs becomes $\delta '$ where
\begin{equation}
|\tan(\delta'/2)|= \tan^2(\delta/2)~.
\end{equation}
After a few rounds, $\delta$ is small --- Alice and Bob have extracted from the initial supply of unknown entangled states a smaller number of known entangled states.

Now this sounds like it could be useful. If Alice and Bob perform QCS using the pairs with a systematic phase error, that systematic error will show up in their measurement of their time offset. If they can trade in the original pairs with unknown phase for a reduced supply of pairs with known phase, it's a win. But as already discussed, the ``purification'' protocol actually replaces the original supply with a reduced supply where the phase of the new pairs is determined by their time offset and therefore {\em still} unknown. Alice and Bob are no better off.

The ``recurrence'' protocol \cite{bdsw} that we have described is relatively simple to execute (though still not easy), but if we are willing to do more sophisticated coherent processing, there are much more efficient protocols that waste far fewer pairs. In particular, there is a ``hashing'' protocol \cite{bdsw}, requiring only one-way communication from Alice to Bob, that (according to standard Shannon arguments) yields, from $n$ initial pairs with fidelity $F$, a supply of distilled pairs, where the fidelity of the distilled pairs is as close as desired to $1$; the number $n'$ of distilled pairs is  asymptotically (for large $n$) close to
\begin{equation}
\label{shannon_yield}
n'=n\cdot\left[1-H_2(F)\right]~,
\end{equation}
where $H_2(F)=-F\log_2 F - (1-F)\log_2(1-F)$ is the binary entropy function. (I'm assuming that the only errors we have to worry about are phase errors.)

This hashing protocol really seems to be a quantum protocol: it involves collective measurement of many qubits at once, and I don't think there is an analogous operation that could be performed on a supply of classical analog clocks. Furthermore, since if has a much better yield of highly distilled pairs than the recurrence protocol, it really does seem to be capable, in principle, of significantly improving our sensitivity to the time offset between Alice and Bob.

But there's a problem, really the same problem as that encountered when we imagined that Charlie creates encoded pairs that are actively stabilized through quantum error correction. To approach the optimal yield of distilled pairs given in Eq.~(\ref{shannon_yield}), Alice and Bob use a phase-error correcting quantum code that they have agreed on in advance. Alice measures the stabilizer generators of the code and sends the measurement outcomes to Bob, who also measures the same generators and then corrects errors to prepare a supply of high fidelity encoded pairs. But the pairs are encoded, and the natural evolution of the qubits does not preserve the code space. Therefore, Alice and Bob need to decode the pairs before performing QCS, and once again they will only be able to detect the offset in QCS {\em relative} to the offset used in the decoding (and the measurement of the stabilizer operators).

In short, if quantum information really offers an advantage for clock synchronization, we don't seem to be realizing that advantage in the QCS protocol, as far as I can see. Perhaps it is because the protocol is really ``too classical.''

\section{Other quantum protocols with a time offset}

While there seems to be an obstacle to using entanglement purification to improve the reliability of clock synchronization, Alice and Bob can nevertheless use purification to enhance the efficacy of other protocols, such as EPR key distribution or teleportation, even if they do not have synchronized clocks. We have seen that if Bob's clock lags behind Alice's by the unknown offset $\Delta$, then by executing the usual distillation procedure, Alice and Bob can prepare high fidelity pairs in the state
\begin{equation}
|\psi^-(\Delta)\rangle = I\otimes U_\Delta^{-1}|\psi^-\rangle~.
\end{equation}
If, say, Alice wants to teleport the unknown state $|\psi\rangle$ to Bob, then with good $|\psi^-\rangle$'s, Alice's joint measurement on the unknown state and her member of the entangled pair would prepare in Bob's laboratory on the same time slice one of the states $\sigma|\psi\rangle$, where $\sigma\in\{I,X,Y,Z\}$ is known from Alice's measurement. If the pairs are in the state $|\psi^-(\Delta)\rangle$ instead, then Alice's measurement prepares 
\begin{equation}
U_\Delta^{-1}\sigma |\psi\rangle~.
\end{equation}
This state evolves in time $\Delta$ to $\sigma|\psi\rangle$. Therefore, if Bob's operation lags behind Alice's by the same amount in both the purification and the teleportation, the teleportation works normally.

This is good news, because it means that Alice and Bob don't need to know the offset $\Delta$ in order to use purification to improve such protocols. But the bad news is that (consistent with our earlier observations), after purifying  Alice and Bob can't use the fidelity of teleportation as a criterion for judging how well their operations are synchronized.

\section{Observables in relativistic quantum theory}

Let's return now to the theme with which I began: that this workshop invites us to reconsider some aspects of quantum information theory in a relativistic setting. In fact, there is a question about relativistic quantum theory that has bothered me for a while, and ruminating about clock synchronization has stimulated me to reconsider it.

The question is: {\em what is an observable?}. The standard answer in quantum field theory is that an observable is a self-adjoint operator that can be defined on a spacelike slice through spacetime. But this is not the right answer in general, not if we mean by an observable something that could really be measured in principle. For many self-adjoint operators, if they could really be ``measured,'' the measurement would allow spacelike-separated parties to communicate.

When I speak of a ``measurement'' of observable $A$ occuring on a time-slice, I don't mean that the outcome of the measurement is instantly known by anyone. Rather I mean that the density operator $\rho$ decoheres on that time slice as \begin{equation}
\label{meas_so}
\rho\to \sum_a E_a\rho E_a~,
\end{equation}
where $\{E_a\}$ is the set of orthogonal projectors onto the eigenspaces of the observable.
Later on, if information from various locations on the slice arrives at a central location, the outcome can be inferred and recorded. 

Now suppose that Alice and Bob share a quantum state $\rho_{AB}$. At time $t=-\epsilon$, Alice performs a unitary transformation, $U_A$,
\begin{equation}
\rho \to (U_A\otimes I)\rho (U_A^\dagger\otimes I)~,
\end{equation}
at time $t=0$ the superoperator Eq.~(\ref{meas_so}) acts on the state, and at time $t=\epsilon$, Bob performs a measurement on his density operator,
\begin{equation}
\rho_B={\rm tr}_A ~\left[ \sum_a E_a(U_A\otimes I)\rho_{AB} (U_A^\dagger\otimes I)E_a\right]~.
\end{equation}
If Alice and Bob are spacelike separated, then the superoperator can be physically realizable only if it is {\em causal} -- Bob's density operator must not depend on the unitary transformation that Alice applies. 

How can the causal observables be characterized? Are all causal observables physically implementable in principle?

\begin{figure}
\begin{center}
\leavevmode
\epsfxsize=3in
\epsfbox{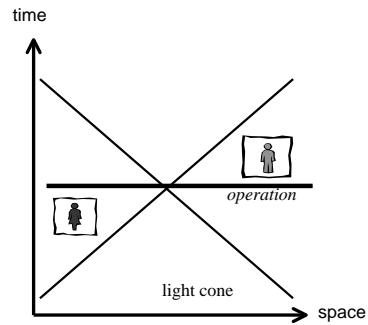}
\end{center}
\caption{An operation on a time slice. If the operation allows spacelike-separated Alice and Bob to communicate, then it is not {\em causal} and hence not physically implementable.}
\label{fig:causal}
\end{figure}

To clarify the concept, let's consider an example, noted by Sorkin \cite{sorkin}, of a measurement that is not causal. It is a two-outcome incomplete Bell measurement performed on a pair of qubits. The orthogonal projectors corresponding to the two outcomes are
\begin{eqnarray}
E_1&=&|\psi^-\rangle\langle \psi^-|~,\nonumber\\
E_2&=&I-|\psi^-\rangle\langle \psi^-|
\end{eqnarray}
Suppose that the initial pure state shared by Alice and Bob is $|00\rangle_{AB}$. This state is orthogonal to $|\psi^-\rangle$, so that outcome 2 occurs with probability one, and the state is unmodified by the superoperator. Afterwards Bob still has a density operator $\rho_B=|0\rangle\langle 0|$.

But what if, before the superoperator acts, Alice performs a unitary that rotates the state to $|10\rangle_{AB}$? Since this is an equally weighted superposition of $|\psi^-\rangle$ and $|\psi^+\rangle$, the two outcomes occur equiprobably, and in either case the final state is maximally entangled, so that Bob's density operator afterwards is $\rho_B=I/2$. Bob can make a measurement that has a good chance of distinguishing the density operators $|0\rangle\langle 0|$ and $I/2$, so that Bob can decipher a message sent by Alice. The measurement superoperator is acausal.

An obvious example of an operation that {\em is} causal is measurement of a tensor product observable $A\otimes B$ -- Alice and Bob can induce decoherence in the basis of eigenstates of a tensor product through only local actions. But there are other examples of causal operations that are a bit less obvious. One is complete Bell measurement, {\it i.e.} decoherence in the Bell basis $\{|\psi^\pm\rangle, |\phi^\pm\rangle\}$. No matter what Alice does, the shared state after Bell measurement is maximally entangled, so that Bob always has $\rho_B=I/2$, and he can't extract any information about Alice's activities. Though Bell measurement is a causal operation, it is not something that Alice and Bob can achieve locally without additional resources.

Now one wonders, why should there exist causal operations that cannot be implemented locally? A possible answer is that there are previously prepared resources that Alice and Bob might share, that while two weak to allow faster-than-light communication, are strong enough to enable decoherence in the Bell basis. One such resource might be shared entanglement; in the case of complete Bell measurement, an even weaker resource will do -- shared randomness.

Suppose that Alice and Bob both have the same string of random bits. This is a useful resource. In particular, the whole idea of quantum key distribution is to establish secure shared randomness -- Alice uses a random key for encoding, Bob for decoding, and if Eve doesn't know the key she can't decode the message. The shared random string also allows Alice and Bob to induce decoherence in the Bell basis. They share a pair of qubits, and on the same time slice, they both consult two bits of the string; depending on whether they read $00,01,10,$ or $11$, they both apply the unitary operator $I,X,Z,$ or $Y$. The superoperator
\begin{eqnarray}
\rho\to && {1\over 4}[(I\otimes I)\rho (I\otimes I) + (X\otimes X)\rho (X\otimes X) \nonumber\\
&+& (Y\otimes Y)\rho (Y\otimes Y) + (Z\otimes Z)\rho (Z\otimes Z)]
\end{eqnarray}
annihilates all the terms in $\rho$ that are off the diagonal in the Bell basis. By a similar method, randomness shared by $n$ parties enables decoherence in the basis of simultaneous eigenstates of any set of commuting operators, where each operator is a tensor product of Pauli operators.

With Dave Beckman, Daniel Gottesman, and Michael Nielsen, I have been mulling over the problem of characterizing causal operations for several months. At first we guessed that any causal superoperator (one that does not allow Alice to signal Bob or Bob to signal Alice) can be implemented if Alice and Bob share entanglement and perform local operations, but Beckman \cite{beckman} discovered counterexamples. 

All of the cases studied so far are consistent with a modified conjecture, suggested  by David DiVincenzo:

\begin{quote}
{\em Conjecture}: A superoperator that does not allow Alice to send a signal to Bob can be implemented with
\begin{itemize}
\item Local operations by Alice and Bob,
\item One-way quantum communication from Bob to Alice.
\end{itemize}
\end{quote}

\noindent Some special cases of this conjecture have been proved \cite{beckman}, but we have no general proof, and in fact I am far from confident that the conjecture holds in general. Even if it does, the conjecture does not provide a very satisfying way to characterize the causal operations. On the one hand, it {\em is} clear that the two resources listed are too weak to allow Alice to signal Bob. But on the other hand, one-way communication from Bob to Alice is not achievable when Alice and Bob are spacelike separated, so an operation that can be implemented with these resources cannot necessarily be applied on a time slice. In quantum physics, there is a perplexing gap between what is {\em causal} and what is {\em local}.

\section{Concluding comments}
It is a captivating challenge to find ways in which quantum error correction and/or entanglement purification can be invoked to improve the accuracy of clock synchronization or frequency standards. But my efforts so far have not met with much success.

I'd like to comment here on an issue that I didn't mention in my talk at the meeting.
A problem related to the issues I discussed in my talk, but in a sense logically independent, arises if we consider in more detail how an ``$X$ measurement'' is actually performed with real atomic clocks: a $\pi/2$ pulse is applied that rotates $X$ eigenstates to $Z$ eigenstates, and then $Z$ is measured. Thus, if both Alice and Bob are to measure $X$, they need to establish a ``phase lock'' to ensure that they are really using the same convention to define $X$. An approach to this problem was suggested in \cite{jpl}, but criticized in \cite{naval} and \cite{shahriar}.

At the meeting, both Dave Wineland and Paul Kwiat proposed an alternative approach. They suggested applying the concept of a ``decoherence-free subspace'' \cite{dfs} to encode Alice's phase convention in a robust two-qubit stationary state that can be sent to Bob.
This proposal relies on the assumption that two qubits transported together will be subjected to identical phase errors in the preferred basis $\{|0\rangle,|1\rangle\}$; thus an eigenstate of the operator $Z\otimes I + I\otimes Z$ will be invulnerable to phase errors. If we encode the logical qubit $a|0\rangle + b|1\rangle$ as
\begin{equation}
a|01\rangle + b|10\rangle~,
\end{equation}
then the encoded state resists dephasing.

Alice and Bob both have local interrogating oscillators that are used to apply $\pi/2$ pulses. At the time that she calls 0, Alice can apply a pulse that prepares $|0\rangle + e^{i\delta}|1\rangle$, she can encode that state as $|01\rangle + e^{i\delta}|10\rangle$, and she can send that encoded state to Bob. Bob can decode it and measure it at the time he calls $T$, and thereby (if Alice sends many such encoded states), he can lock his phase convention at time $T$ to Alice's convention at time $0$. This shared phase convention might be useful, but in itself it does not solve the problem of synchronizing Alice's clock with Bob's. 

We can also ask how well the assumptions of a preferred dephasing basis, and identical phase errors on qubits traveling together, apply in a realistic physical setting. These may be reasonable assumptions if $|0\rangle$ and $|1\rangle$ are energy eigenstates of a two-level atom, and the dephasing is dominated by fluctuating electromagnetic fields. The same idea works if $|0\rangle$ and $|1\rangle$ are linear polarization states of a photon, aligned with the preferred axes of the optical medium \cite{kwiat}. 

The clock synchronization problem might be viewed as an entry into a fascinating subject --- relativistic quantum information theory. I've reported here on some modest progress toward a better understanding of one issue in this theory: the structure of causal observables.

Observations related to those in this talk appear in a recent paper by Yurtsever and Dowling \cite{yurtsever}. 

\acknowledgments
I thank Jon Dowling for organizing this meeting and for encouraging me to write up this report. I'm also grateful to the participants, especially Xinlan Zhou, for stimulating discussions, and to Dave Wineland and Paul Kwiat for helpful correspondence. My interest in the clock synchronization problem was originally stimulated by discussions with Hideo Mabuchi and Jeff Kimble; flaws in some of my ideas were pointed out by Steven van Enk. The comments about causal operations are based on work with Dave Beckman, Daniel Gottesman, and Michael Nielsen, aided by useful advice from David DiVincenzo. This work has been supported in part by the Department of Energy under Grant No. DE-FG03-92-ER40701, by DARPA through the Quantum Information and
Computation (QUIC) project administered by the Army Research Office under Grant
No. DAAH04-96-1-0386, and by an IBM Faculty Partnership Award.

\end{document}